\documentclass[aps, pre, twocolumn, showpacs]{revtex4}
\usepackage{graphicx}
\usepackage{amsmath}
\usepackage{amssymb}
\usepackage{wasysym}

\begin{document}

 \title{Frequency structure of the nonlinear instability\\of a dragged viscous thread}
\author{Robert L. Welch, Billy Szeto and Stephen W. Morris}
\affiliation{Department of Physics, University of Toronto, 60 St. George St., Toronto, Ontario, Canada M5S 1A7}
\date{\today}

\begin{abstract}
A thread of viscous fluid falling onto a moving surface exhibits a spectacular variety of types of motion as the surface speed and
nozzle height are varied. For modest nozzle heights, four clear regimes are observed.  For large surface speed, the thread is
dragged into a stretched centenary configuration which is confined to a plane.  As the surface speed is lowered, this exhibits
a supercritical bifurcation to a meandering state.  At very low surface speeds, the state resembles the usual coiling
motion of a viscous thread falling on a stationary surface.  In between the meandering and coiling regimes, a window containing a novel
multifrequency state, previously called ``figures of eight" is found.  Using an improved visualization technique and a fully automated apparatus, we made  detailed measurements of the longitudinal and transverse motion of the
thread in all these states.
We found
that the multifrequency state is characterized by a complex pattern of motion whose main frequencies are locked in a 3:2
ratio.  This state appears and disappears with finite amplitude at sharp bifurcations without measurable hysteresis.
\end{abstract}

\pacs{82.40.Bj, 47.20.Gv, 47.20.Ky}

 \maketitle

A thread of viscous fluid falling onto a surface, such as honey falling onto a piece of toast, spontaneously wraps itself into beautiful coils~\cite{barnes_woodcock, barnes_mckenzie,ann_rev_2011}. This ``liquid rope coiling instability" is due to the buckling that arrises from the competition between axial compression and bending as the thread impacts the surface~\cite{taylor}.  Similar buckling instabilities are responsible for folding patterns in sheets of viscous fluid~\cite{maha2} for example in poured paint~\cite{pollock}, flowing cake mix or in certain lava flows~\cite{pahoehoe}.  Modifying the coiling  instability by moving the surface onto which the thread is falling, say by translating the toast uniformly, leads to a rich panoply of new states of motion.  Such a ``fluid mechanical sewing machine" ~\cite{chu_webster, meander_linear_stability,MDRL2008,blount_lister} is in many ways more experimentally controllable than pure liquid rope coiling, which can lead to tall, unstable stacks of coiled material~\cite{barnes_woodcock,barnes_mckenzie, maleki_PRL}.  These stacks create an ill-controlled lower boundary condition.  Moving the surface rapidly drags the thread into a stretched catenary~\cite{chu_webster} which forms an especially simple, high symmetry basic state of steady flow. The bifurcation from the catenary to the single frequency meandering state as the translation speed is lowered has been examined in detail both theoretically~\cite{meander_linear_stability} and experimentally~\cite{MDRL2008}.  The meandering state, in which the motion is a simple transverse oscillation, is in many ways a  reduced symmetry version of the classic liquid rope coiling instability, but  one with a very well controlled lower boundary condition.

In this paper, we extend the experimental study of the bifurcations of a dragged viscous thread using a different, more accurate technique for measuring the position of the thread and more precise control of the experimental parameters.
%
%
We used this improved technique to quantitatively examine the more complex nonlinear states that occur as the speed is lowered further.  To keep things simple, we confine ourselves to the region of nozzle heights below which 
 the higher pendulum modes of the thread remain unexcited.  In this regime, a window of fascinating multi-frequency motion, called ``figures of eight" in Refs~\cite{chu_webster,MDRL2008}, appears between the meandering and translated coiling regimes.  It consists of fluid loops falling on alternate sides of the main thread.  Using an automated apparatus and an improved visualization technique, we explored the region in and around the ``figures of eight" window.  We measured the frequencies, amplitudes and phase relationships between the longitudinal and transverse motions of the thread and used this analysis to objectively categorize the states.  This makes possible a much more detailed study of the state diagram and of the frequency structure of the nonlinearly coupled modes of the motion.  We find, in particular, that despite its name, the ``figures of eight" state contains several modes with an unexpected 3:2 frequency ratio.  The thread executes a 
surprisingly delicate and complex dance over the moving surface  to produce the alternating loop pattern.

\section{experiment}
\label{expt}

\begin{figure}
\includegraphics[height=6.4cm]{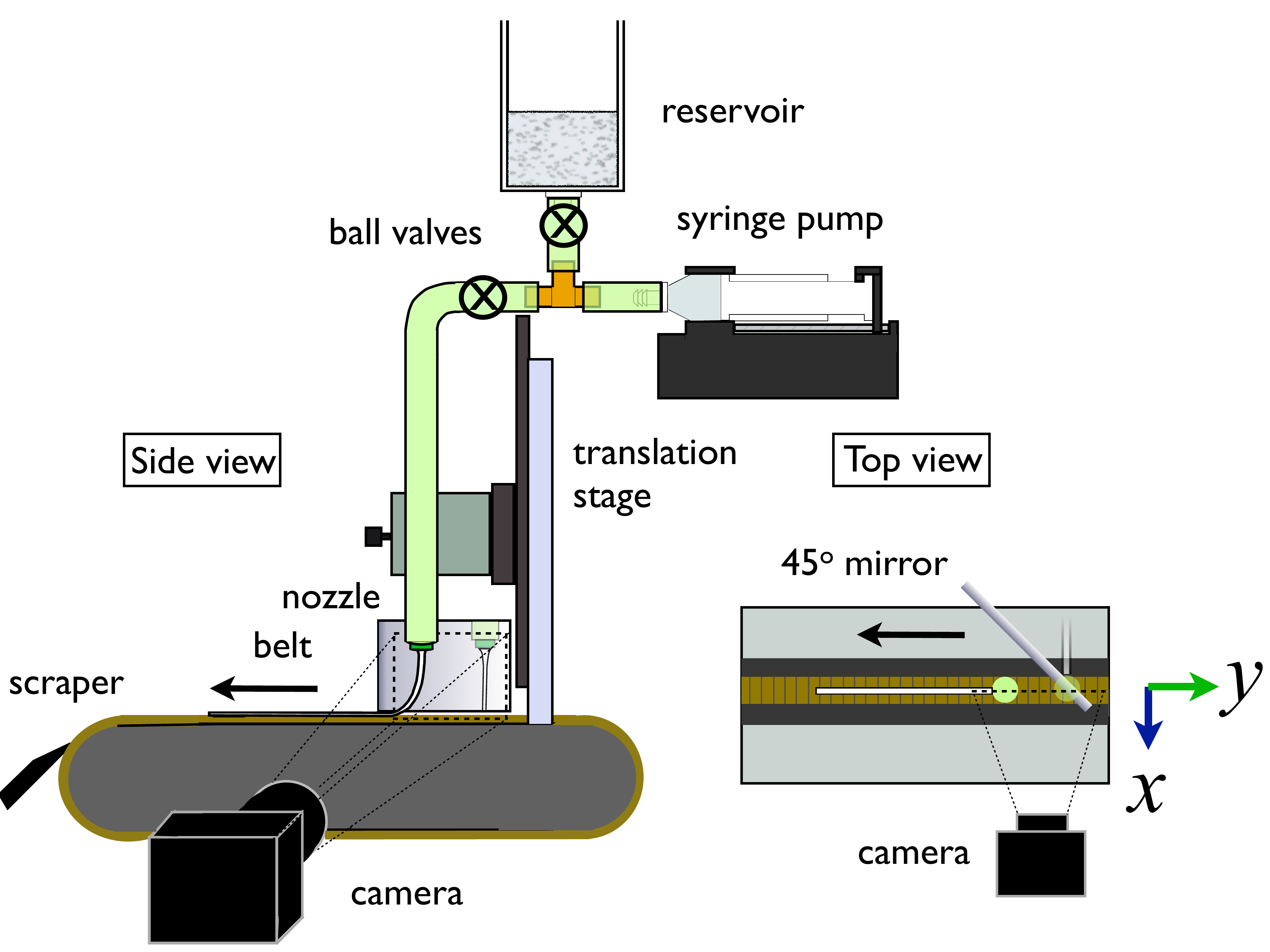}
\caption{(Color online) A schematic of the experimental apparatus, showing side and top views.  The belt moved at speed $U$ in the direction of the arrow. The camera simultaneously collected $x$ and $y$ views by means of a 45$^\circ$ mirror. }
\label{apparatus}
\end{figure}

The apparatus, shown schematically in Fig.~\ref{apparatus}, is an improved version of the one described in Ref.~\cite{MDRL2008}.  A  16~mm wide toothed timing belt formed the moving surface.  The upper surface of the belt was smooth. As the belt passed over the downstream pulley, a plastic scraper removed the accumulated fluid, so that the belt returned carrying only a very thin coating of oil.  The removed fluid was collected and recycled.   The belt speed $U$ was controlled by a very fine  50~000 step/rotation indexed stepper motor whose drive frequency was provided by a computer controlled signal generator.  The motor provided a belt speed which was several orders of magnitude more precise than the DC motor used in Ref.~\cite{MDRL2008}.  The fluid issued from the same stainless steel nozzle used previously, whose diameter was $d = 8.00 \pm 0.02$~mm.  The vertical height $H$ of the nozzle above the belt was determined by screw-driven computer controlled linear translation stage with a full range of 14~cm.  The precision of the stage was $\pm 0.5 \mu$m, although the absolute accuracy of $H$, which was mainly limited by the need to physically locate the $H=0$ nozzle  position, was somewhat less than this.  Here, we consider only the simplest motions, which occur for $H \le 6.0$~cm.

The fluid was identical to that used in Ref.~\cite{MDRL2008}, namely 30\,000~cSt  Dow Corning 200\copyright   ~silicone oil.  This oil is known to be newtonian and its properties are only very weakly temperature dependent.  The relevant fluid properties of the oil are its density $\rho$, kinematic  viscosity $\nu$ and surface tension $\sigma$; these are tabulated in Ref.~\cite{MDRL2008}.  

The volumetric flow rate $Q$ from the nozzle was set by a computer controlled syringe pump.  The syringe could be periodically refilled from a reservoir {\it via} a valve system, as shown in Fig.~\ref{apparatus}.  Using an infuse rate of 1.81 mL/min, we produced a measured mass flux rate of $\rho Q = 0.0270 \pm 0.0008$~g/s.  This is, within uncertainty, the same flux rate as was studied in Ref.~\cite{MDRL2008}, which used a gravity feed system rather than a syringe pump.  Although the syringe pump allows computer controlled variation of $\rho Q$, which may lead to some interesting effects~\cite{ribe_huppert_JFM}, in this paper we kept this quantity fixed. 

 The various physical properties can be collected into several dimensionless groups, as discussed in Ref.~\cite{MDRL2008}.  By restricting $H$ to be less than 6~cm, the present version of the experiment covered a smaller range of states than that described in Ref.~\cite{MDRL2008};  all of the states that we discuss below are in the gravitational (G) regime for pure viscous coiling~\cite{ribe_PRSL,ribe_huppert_JFM,coiling_stability}.

\begin{figure}
\includegraphics[width=7cm]{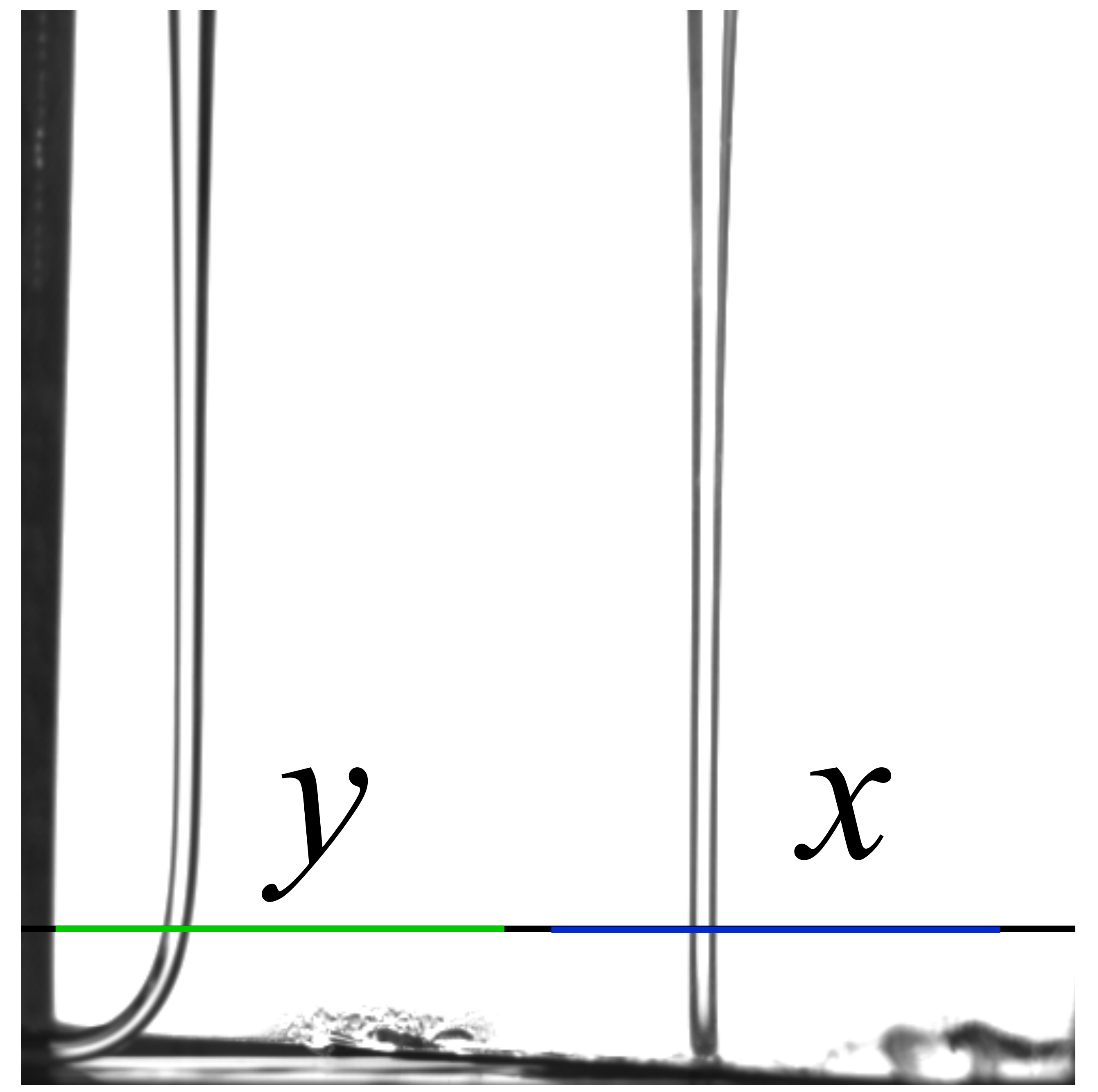}
\caption{(Color online) The two simultaneous views of the thread, as seen by the camera.  The belt moves to the left with speed $U$. The righthand image is the view of the thread transverse to the belt motion, seen in the mirror, while the lefthand one is the longitudinal view, seen directly.  The horizontal line indicates the portion of the image used to determine the $x$ and $y$ positions of the thread.  The thread is shown in the catenary state just above the meandering transition. }
\label{edgedetect}
\end{figure}

In previous studies~\cite{chu_webster,MDRL2008}, the pattern of motion of the thread was visualized by a camera looking down onto the belt.  Here, we use a side-view visualization scheme similar to the one shown in preliminary form in Fig.~9 of Ref.~\cite{MDRL2008}.  As shown in Figs.~\ref{apparatus} and \ref{edgedetect}, a computer-controlled camera simultaneously imaged the viscous thread in both a side view ({\it i.e.} in the longitudinal plane containing the midline of the belt) and a front view ({\it i.e.} a transverse plane across the width of the belt).  The latter view was provided by a 45$^\circ$ mirror positioned upstream of the contact point of the thread, as shown in Fig.~\ref{apparatus}.  The overall position and shape of the thread was thus imaged in both the $x$ (transverse) and $y$ (longitudinal) directions, as a function of time.  

To image the thread with good contrast, a cool white flat panel fibre optic light source was positioned off the downstream end of the belt so that it formed a bright background for both views in the mirror.  The edges of the transparent thread then show as sharp dark features, as seen in Fig.~\ref{edgedetect}. 

 For nozzle heights $H \le 6.0$~cm, the overall shape of the thread consists of a smooth curve with a single maximum in the $y$  direction --- the maximum being the upstream position of the ``heel" that is formed by the buckled end of the thread a few mm above the surface of the belt~\cite{blount_lister}.  We selected a horizontal line from each image, as shown in  Fig.~\ref{edgedetect}, and used image analysis along this line to locate the $x$ and $y$ positions of the thread.  The centerline of the thread was determined with sub-pixel resolution by fitting to a model of the greyscale expected for a transparent cylinder. These centerline positions are not identical to the actual contact point of the thread with the belt, but they are proportional to them.  From these positions, the $x$ and $y$ motion of the contact point can be reconstructed~\cite{shape_note}.   
 
 This imaging scheme has several advantages over imaging the pattern by viewing the belt from above as was done previously~\cite{chu_webster,MDRL2008}.  The $x$ and $y$ motions are easily separated and the contrast of the side-view images is uncluttered by the mesh textures and lighting problems associated with imaging the belt through the transparent thread of fluid lying on top of it ({\it c.f.} Fig.~2 of Ref.~\cite{MDRL2008}).  In the following, we deduce and classify the state of motion of the thread, as a function of $H$ and $U$, using 150~s time series of side view images taken at 25 frames per second.

\begin{widetext}
\begin{center}
\begin{figure}
\includegraphics[width=14cm]{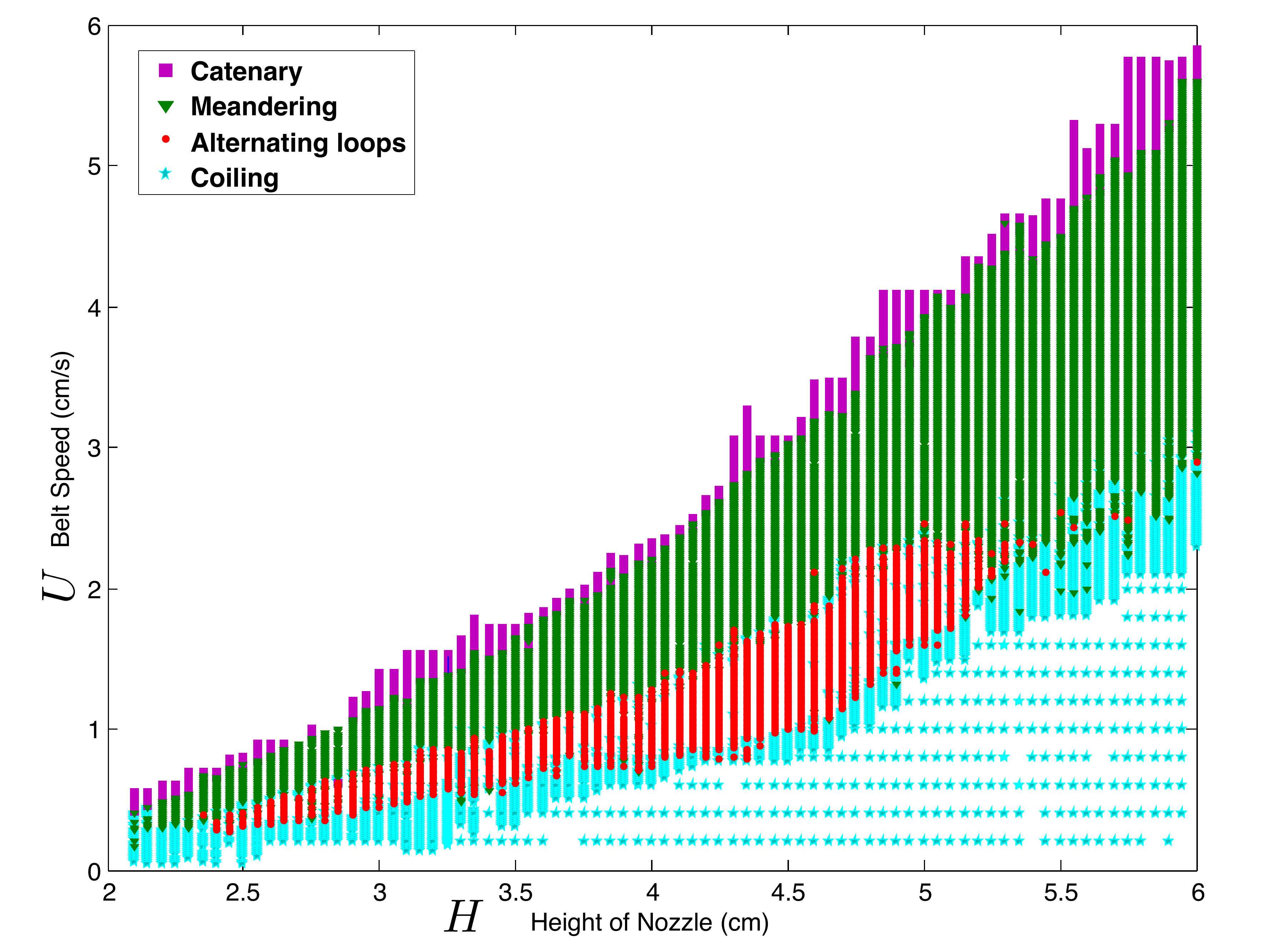}
\caption{(Color online) The state diagram, obtained by automated variation of $H$ and $U$, with automated state classification based on the Fourier spectra of the $x$ and $y$ motions.  The window of the alternating loop state (small circles)
is bounded on all sides by the meandering (triangles)
and coiling (stars)
states.  The onset of fluid pendulum behavior occurs for $H \ge 6$~cm, just beyond the righthand edge of this plot.}
\label{phasediagram}
\end{figure}
\end{center}


\begin{center}
\begin{figure}
\includegraphics[width=8cm]{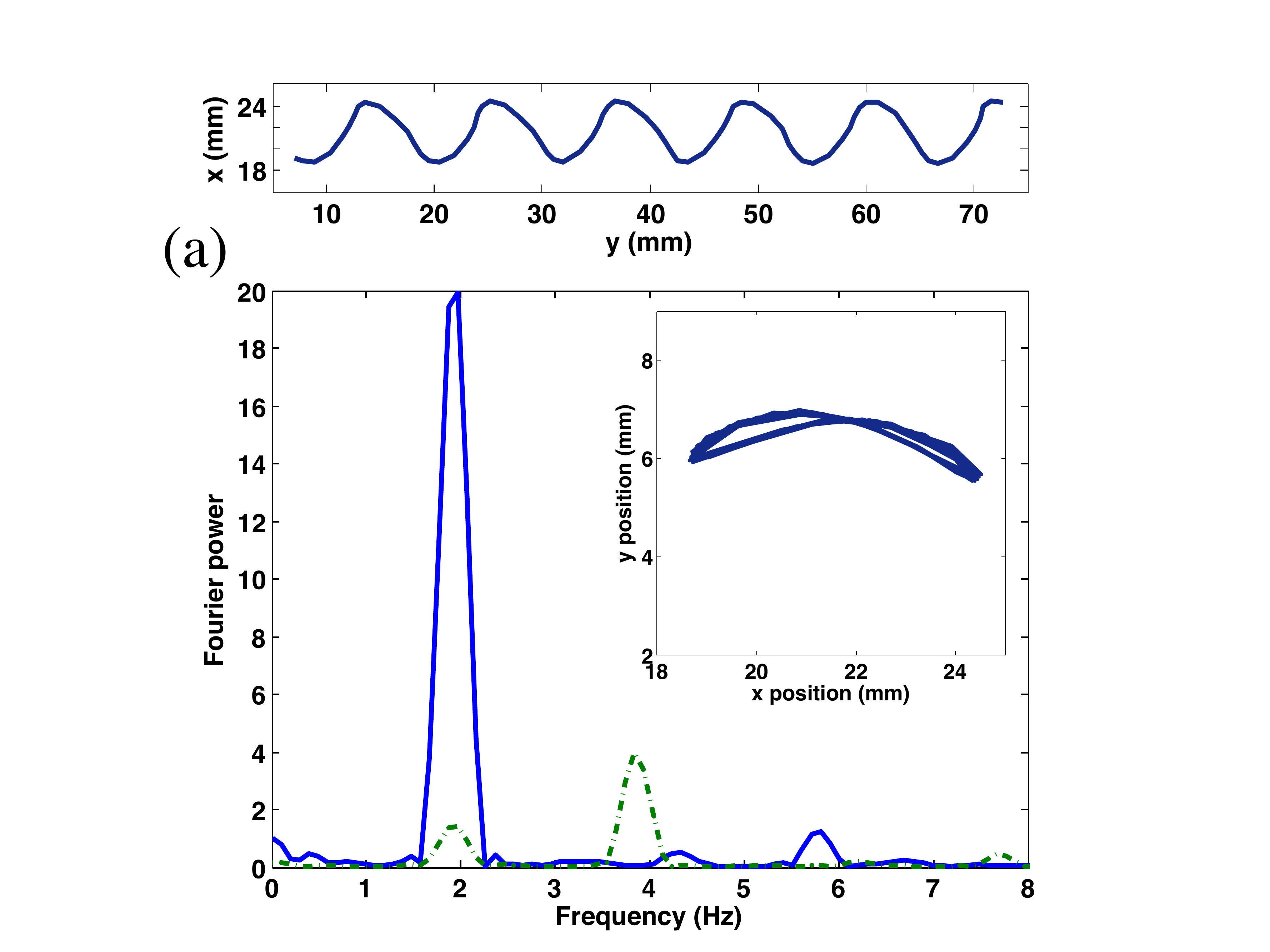}
\includegraphics[width=8cm]{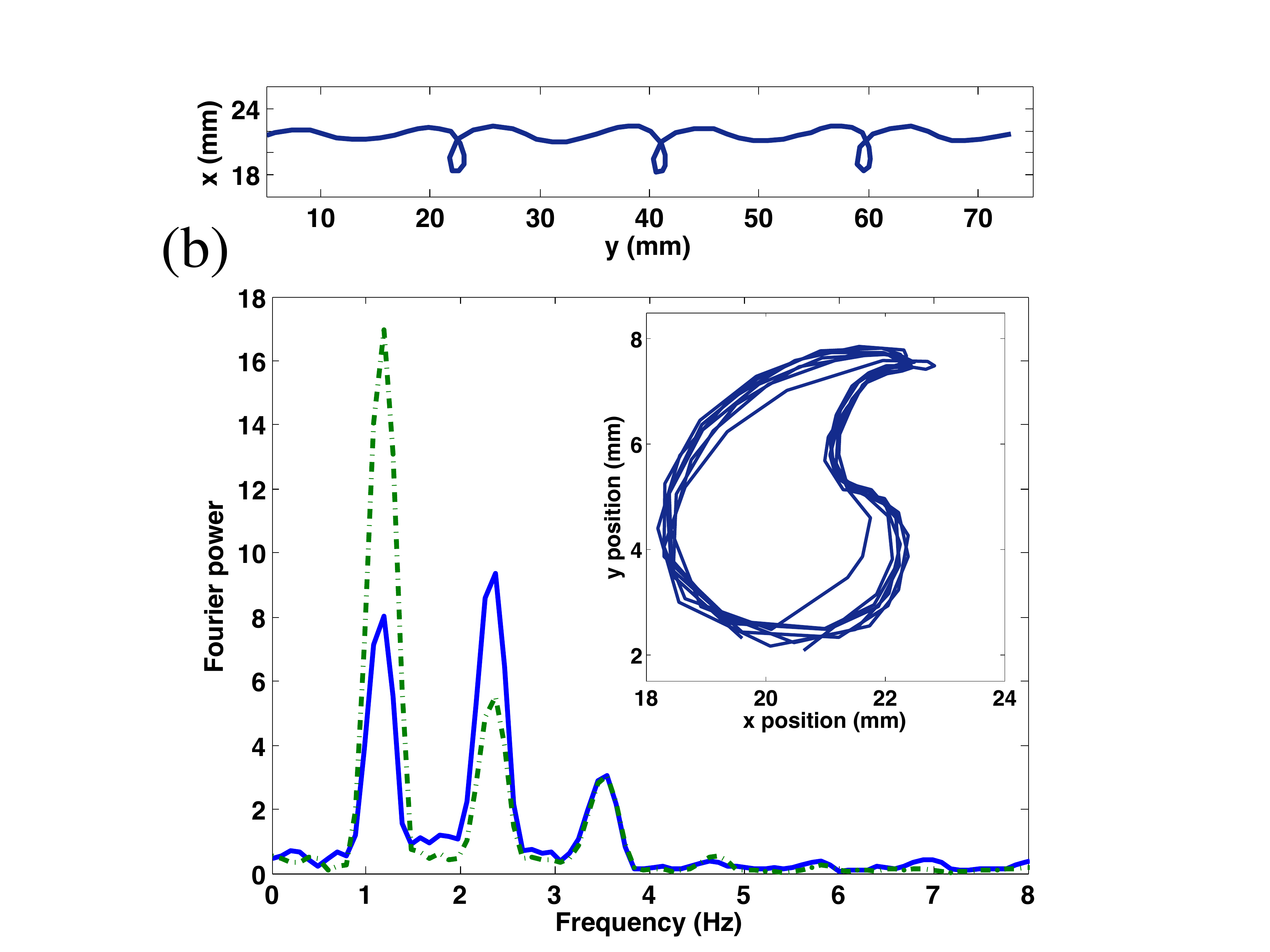}
\includegraphics[width=8cm]{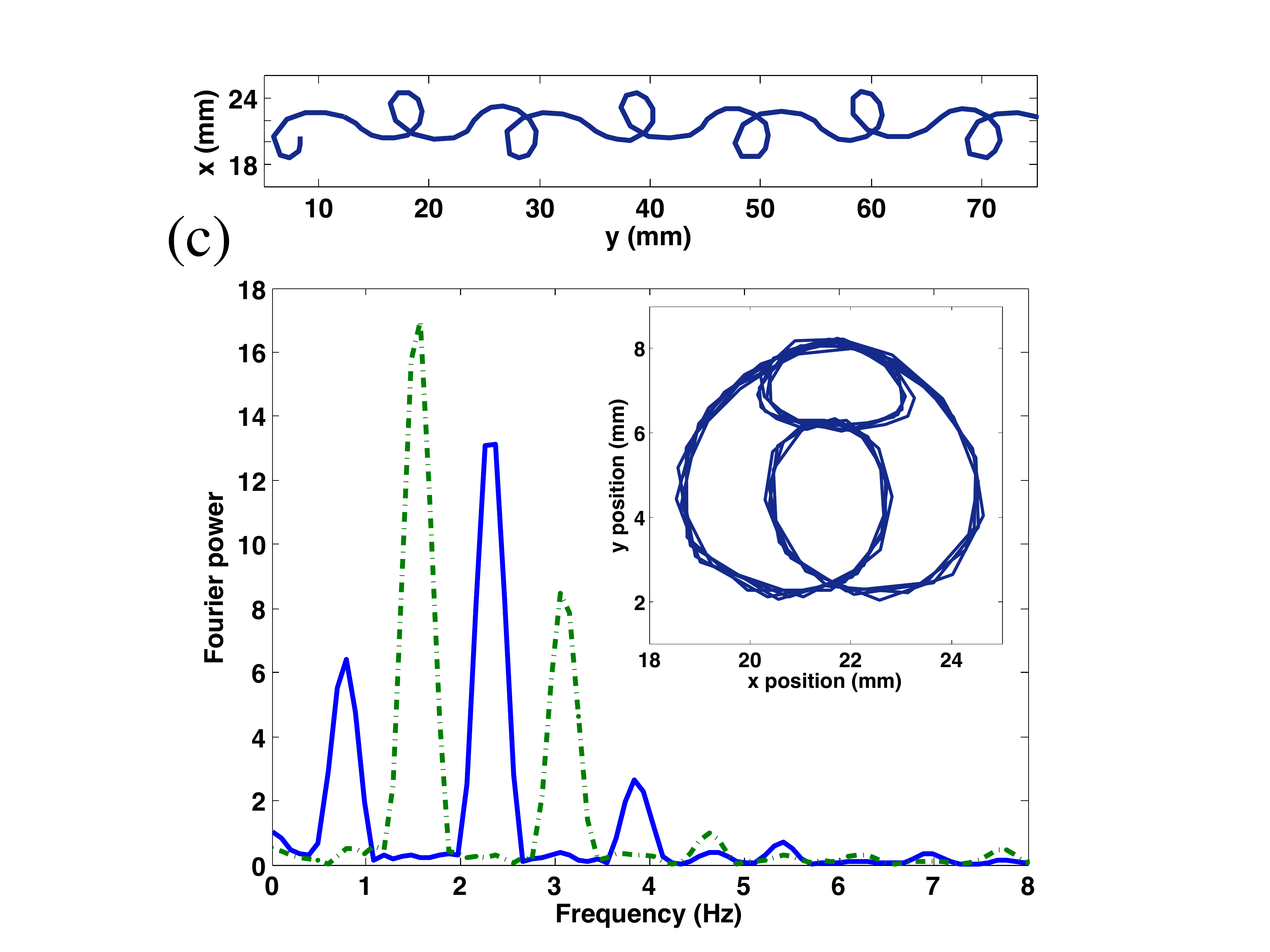}
\includegraphics[width=8cm]{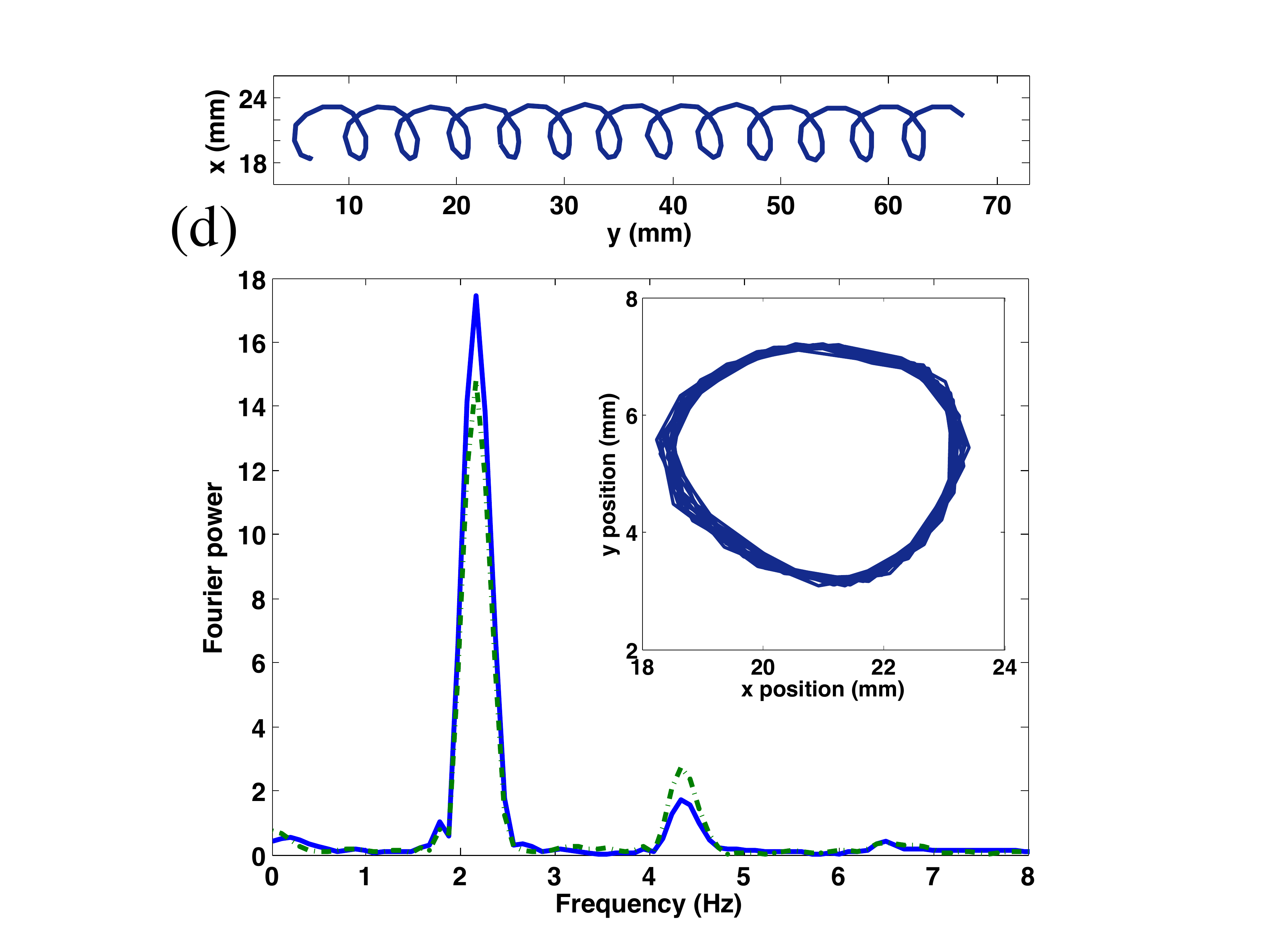}
\caption{(Color online) The Fourier spectra of the $x$ (
solid) and $y$ (
dashed) motions for the (a) meandering ($U = 2.23$~cm/s), (b) one side looping ($U = 2.19$~cm/s), (c) alternating loop ($U = 1.60$~cm/s) and (d) coiling states ($U = 1.00$~cm/s).  Each panel shows the power spectra, with a reconstruction of the pattern traced by the tip of the thread on the belt above and an inset showing the corresponding Lissajous figure.  For all these data,  $H = 5.00$~cm. For the Lissajous figures, the upstream belt direction is at the top.}
\label{spectra}
\end{figure}	
\end{center}
\end{widetext}

\section{results and discussion}
\label{results}


The automated apparatus allows a systematic exploration of the parameter space spanned by the nozzle height $H$ and the belt speed $U$.  As in Ref.~\cite{MDRL2008}, we left the other parameters constant and examined the states of motion as $U$ was increased and decreased, for fixed $H$.  At each $H$ and $U$, a time series of images was taken after a brief settling time.  Fig.~\ref{phasediagram} shows the the resulting phase diagram.  

The various states of the motion of the thread were automatically characterized by a Fourier technique described below.  This automated procedure makes it possible to obtain orders of magnitude more data than was available in previous studies~\cite{chu_webster,MDRL2008}. Each data point in Fig.~\ref{phasediagram} is the result of a quantitative, objective measurement, rather than a state assigned by an ``eyeball" classification based on scrutinizing images~\cite{chu_webster,MDRL2008}.

As has been observed previously, four distinct types of motion exist over wide areas of the parameter space for $H \le 6$~cm.  At high belt speed, the catenary base state is found, in which the thread remains in a vertical plane.  As the belt speed is lowered, this bifurcates into a meandering pattern.  This non-hysteretic Hopf bifurcation~\cite{meander_linear_stability} was studied in detail in Ref.~\cite{MDRL2008}.

At the lowest belt speed, the thread executes a translated version of the coiling state found for zero belt speed.  In between meandering and translated coiling, there exists an intriguing state which was called the ``figure of eight" state in previous work~\cite{chu_webster,MDRL2008}, but which is better termed the ``alternating loop" state for reasons that will become clear below.  This state exists in a window which is bounded above and below by the meandering and translated coiling states. At the small $H$ end of the window, it appears to connect to a direct transition between meandering and coiling, although the very thick thread and slow dynamics in that region makes precise definitions difficult.  At its high $H$ end, the alternating loop window adjoins a region of hysteretic transitions between meandering and coiling.  The transitions between the  alternating loop and the meandering and coiling states are direct and not measurably hysteretic.


The most straightforward way to characterize the tip motion in the lab frame, given by the two time series $x(t)$ and $y(t)$,  is to examine their Fourier spectra.  The spectra associated with four different states are shown in Fig.~\ref{spectra}, for a traverse of the state diagram shown in Fig.~\ref{phasediagram} corresponding to the  vertical line $H=5.0$~cm. The trivial catenary state is not shown. In addition to meandering, alternating loops and coiling, we occasionally found one side looping, or ``kidney bean" states, but these motions did not appear over any significant area of the state diagram.  They tended to occur near the boundaries of the alternating loop window.  

For each of the $x(t)$ and $y(t)$ time series, the corresponding Fourier spectra, calculated with a Hanning prefilter, are shown in Fig.~\ref{spectra}.  We can also reconstruct the tip motion in the reference frame of the belt, and thus the pattern of fluid left on the belt, by plotting $(x(t), y(t) - Ut)$, using the measured belt speed $U$.   These belt-frame trajectories are also shown in Fig.~\ref{spectra} and resemble the top-view images of the pattern on the belt which were shown in Ref.~\cite{MDRL2008}.  Unlike top-view images, which in Ref.~\cite{MDRL2008} were manually analyzed in the meandering state only, the $x(t)$ and $y(t)$ time series data are quantitative measurements of the thread motion that can be analyzed to extract amplitudes and frequencies of motion in all of the different states the system presents, not just meandering.  The $x(t)$ and $y(t)$ time series contain information which would be very difficult to extract from top-view images of the fluid left on the belt.
%
%

Finally, the Lissajous figure~\cite{lissajous} traced by the tip of the thread, $(x(t), y(t))$ is  shown in the insets of Fig.~\ref{spectra}.  Analyses of these kind were automatically carried out for every point shown in the state diagram in Fig.~\ref{phasediagram}.  The distinct features of the Fourier spectra, discussed below, were used to automatically determine the state plotted in the state diagram. 

The Fourier spectra of the various states show clearly the striking simplicity of the motions.  The meandering state shows the obvious cross-belt $x$ oscillation at a frequency related to the Hopf frequency $\omega$ of the bifurcation~\cite{MDRL2008} locked to a $2\omega$ longitudinal $y$ motion. The nonlinear origin of this 2:1 frequency locking was explained in terms of a simple model in Ref.~\cite{MDRL2008}.  The nonlinearity of the oscillation is manifest in the slight bending distortion of the corresponding Lissajous figure.  Similarly, the spectra of the translated coiling state simply consist of a single frequency oscillation in the $x$ direction locked to a $\pi/2$ out of phase mode with the same frequency in the $y$ direction.  Its Lissajous figure is  nearly a circle, as one would expect. 

The most interesting and surprising Fourier spectra are those of the alternating side looping and of the rarer one side looping states.  The Fourier spectra of the alternating side looping state reveals a multimode motion whose dominant frequencies are locked in a 3:2 ratio.  The Lissajous figure shows an unexpectedly complex structure (which we colloquially call ``the Toyota state").  The Lissajous figure is reflection symmetric in the $x$, or cross-belt, direction, but not symmetric in the $y$, or longitudinal, direction.  Its tighter upper loop, which describes the rapid tip motion while the left and right alternating loops are being executed, is at the upstream end of the Lissajous figure.


\begin{figure}
\includegraphics[width=9cm]{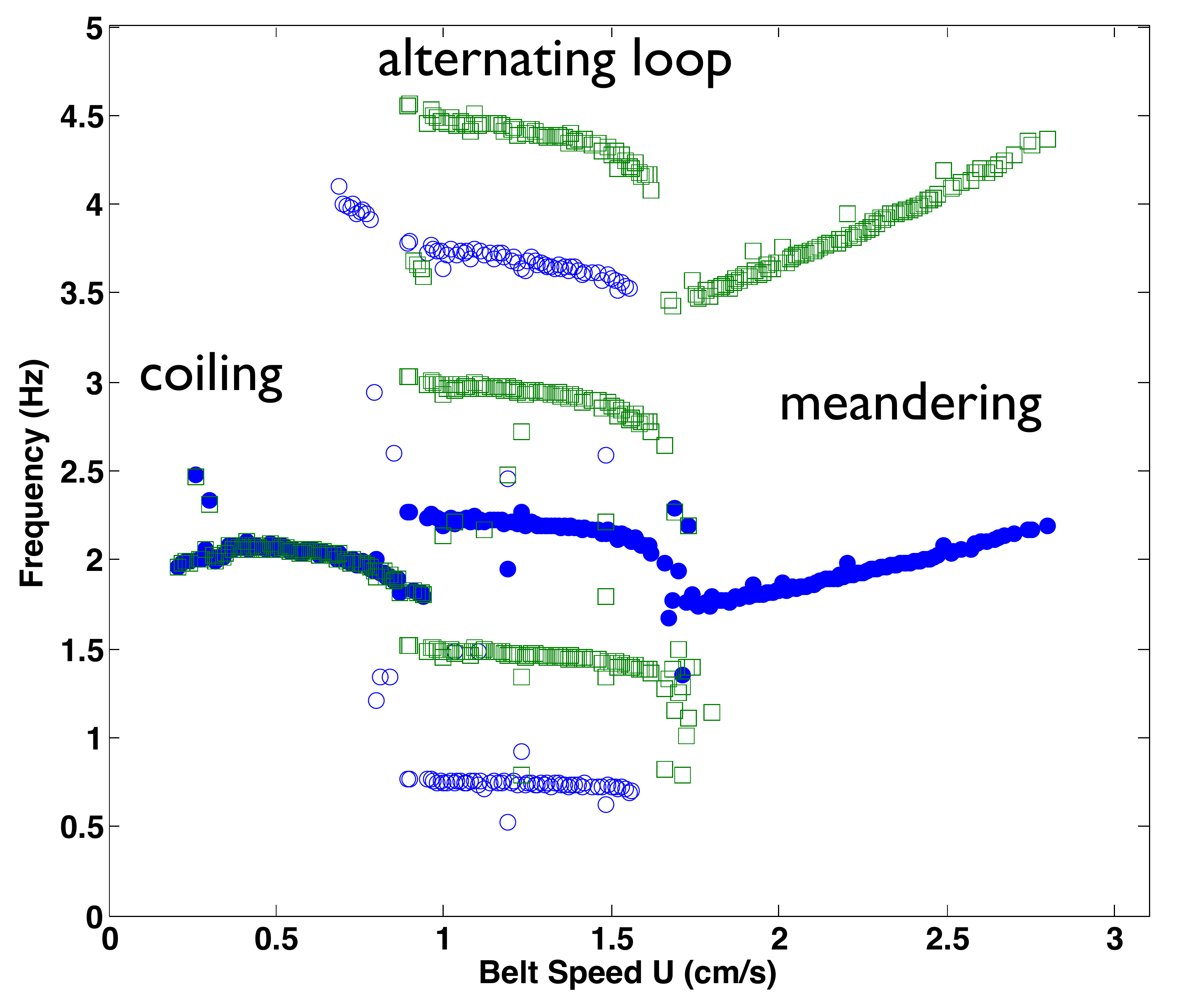}
\caption{(Color online) The frequencies of the main modes of the $x$ (circles) and $y$ (squares) motions of the thread as a function of the the belt speed $U$, for $H = 4.45$~cm. The solid symbols trace the $U$ evolution of the mode corresponding to the main Hopf frequency of the meandering state.  The $x$ and $y$ symbols overlap in the coiling state. }
\label{freqvsU}
\end{figure}

%
%

\begin{figure}
\includegraphics[height=7.6cm]{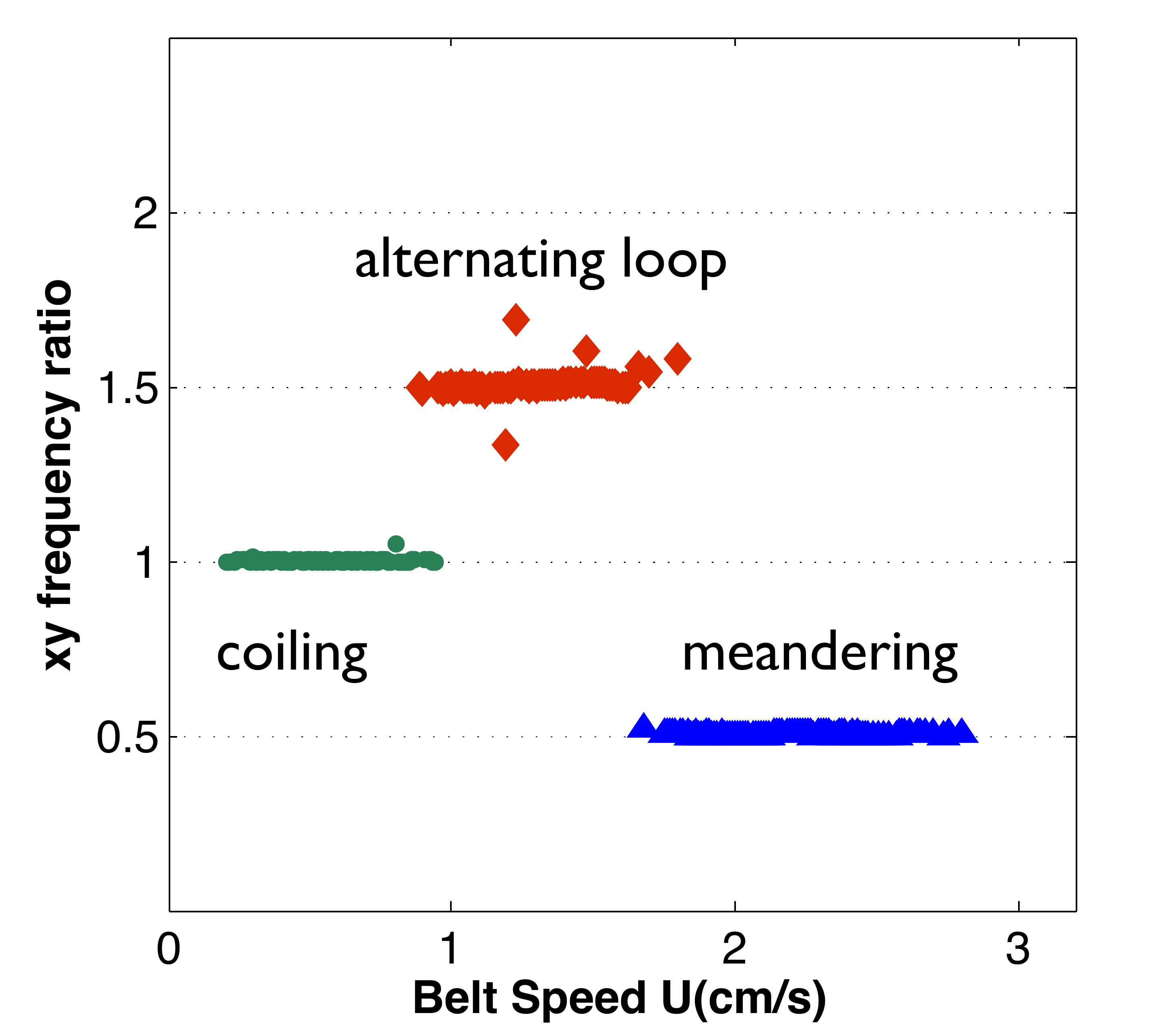}
\caption{(Color online) The ratio of the dominant frequencies of the main $x$ and $y$ modes, {\it vs.} $U$, for $H = 4.45$~cm, using the same data as Fig.~\ref{freqvsU}. This shows clearly the 3:2 nature of the alternating loop state.}
\label{freqratio}
\end{figure}

The Lissajous figure for the one side looping, or ``kidney bean", state shows a distorted circle with a sharp reversal which 
resembles an $x$ reflection symmetry broken version of the Lissajous figure for the alternating side looping state.  The Fourier spectrum of the one side looping state shows a 2:1 frequency locking, with a mixture of mode amplitudes and phase relationships that is quite different from the either the meandering or the one side looping states.  The one side looping state, which is only found in isolated parameter regions,  may be observed in either left handed or right handed versions.



To determine the state plotted on the diagram shown in Fig.~\ref{phasediagram}, we employed a peak finding algorithm on the Fourier spectra of the $x$ and $y$ motions and classified the state according to the frequency ratios of the largest peaks. Occasional ambiguous time series were classified as ``disordered".  Fig.~\ref{freqvsU} shows the frequencies of all the major spectral peaks as a function of the belt speed $U$ for fixed $H = 4.45$~cm.  It can be observed that no peak frequency remains continuous throughout the variation of $U$: small jumps in frequency occur at each state boundary where additional new modes appear.

The meandering state is characterized by a dominant $x$ mode which appears at the Hopf frequency and thereafter decreases linearly with $(U_c-U)$, as discussed in Ref.~\cite{MDRL2008}.  It is accompanied by a frequency doubled $y$ motion, as discussed above. Within the alternating side looping window, the $x$ and $y$ motions each have three dominant frequencies separated by equal frequency increments.  The main $x$ mode in the meandering regime experiences a small upward jump as the alternating side looping window is entered from above and new $x$ modes appear at 1/3 and 5/3 of the dominant frequency.  Meanwhile, $y$ motions appear at 2/3, 4/3 and twice the frequency of the dominant $x$ mode.  As the belt speed is reduced into the translated coiling regime, the extra peaks disappear and the $x$ and $y$ modes lock to one frequency with a small downward jump.  As the $U=0$ coiling state is approached, the single dominant frequency returns to very close to the Hopf frequency at which meandering first appears.  This agrees with previous calculations and observations~\cite{meander_linear_stability,MDRL2008}  which noted the near coincidence of these two frequencies.  

 Fig.~\ref{freqratio} shows the ratios of the frequencies of the dominant peaks in the spectra as a function of the belt speed $U$.   It is evident that the alternating looping state, which was formerly known as the ``figure of eight" state, was poorly named, since its frequency content indicates that it is nothing like the 2:1 resonance envisioned in Ref.~\cite{MDRL2008}.  This nomenclature was based on visual observations only.   In fact, it is the meandering state which exhibits the only 2:1 frequency ratio, while the alternating loop state very clearly involves a 3:2 locking phenomenon.  The $x$ and $y$ motions in the translated coiling state are of course locked in a 1:1 ratio.


\begin{figure}
\includegraphics[width=8.5cm]{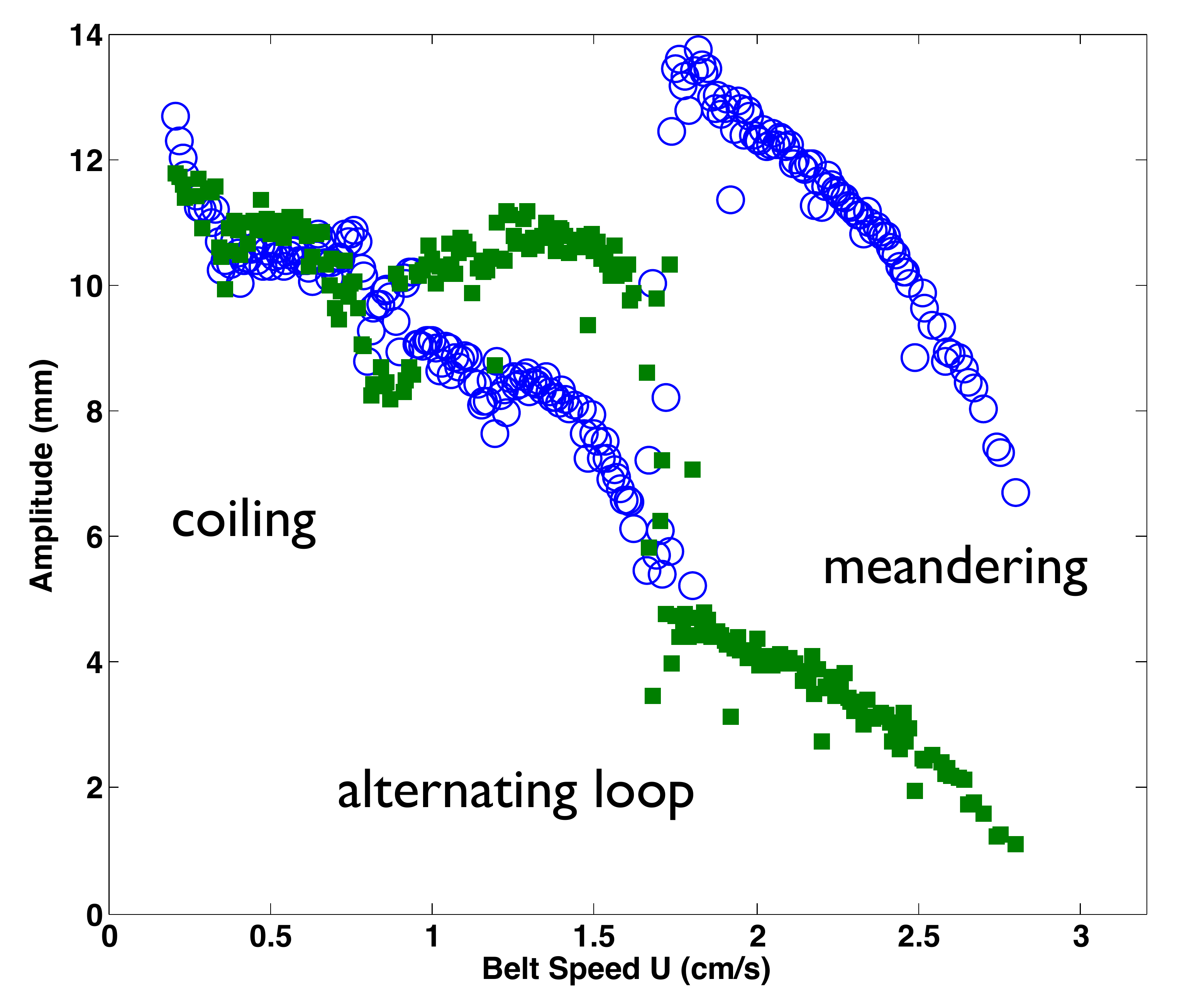}
\caption{(Color online) The amplitudes of the  largest $x$ (open circles) and $y$ ( solid squares) Fourier modes as as a function of $U$, for $H = 4.45$~cm, using the same data as the previous two Figures.}
\label{ampvsU}
\end{figure}

Fig.~\ref{ampvsU} shows the amplitudes of the largest Fourier modes as a function of $U$ for fixed $H$, using the same data plotted in the previous figure.  
The increase in the amplitude of the $x$ mode as $U$ is decreased below the onset of meandering was analyzed in detail in Ref.~\cite{MDRL2008}, where it was shown that the amplitude is well described by the square-root dependence expected for a forward Hopf bifurcation~\cite{meander_linear_stability}.
Upon the transition into the alternating loop window from meandering as $U$ is further decreased,  the amplitude of the $x$ motion exhibits a sharp drop while the amplitude of the $y$ motion increases.  The thread behaves as if it is almost inextensible because the degree of stretching is limited.  Evidently, the amplitudes must adjust themselves in order to provide the required arc length without too much stretching as the alternating loops are traced out.    Finally, the amplitudes become nearly equal in the translated coiling regime, which resembles the symmetric circular coiling found at $U=0$.

\section{conclusions}
\label{conc}

We have constructed an automated apparatus to explore the complex motion of a viscous thread falling onto a moving surface.  Holding the nozzle diameter, the volumetric flow rate and fluid parameters  constant, we varied the nozzle height $H$ and belt speed $U$ for the lower range of speeds and heights $H \le 6.0$~cm.  This corresponds to the gravitational regime~\cite{ribe_PRSL,ribe_huppert_JFM,coiling_stability}  below the onset of multifrequency behavior where the pendulum motions of the thread begin to interfere with the basic coiling frequency.  In this region, the states are relatively simple and mostly non-hysteretic and the thread takes on a monotonic shape.  Using a side-view visualization method, we measured the position of the tip of the thread in the planes transverse and longitudinal to the motion of the belt.  Using a Fourier analysis of these two motions, we were able to automatically classify the states of the thread and make a more detailed state diagram based on objective definitions of the states.  

We mapped the catenary, meandering, alternating loop and translated coiling states, as well as observing a novel broken symmetry one side looping state that was only found in isolated regions of the parameter space.  The origin of the symmetry-breaking and the relationship of this state to the nearby alternating loop state is not obvious. Although their Lissajous figures appear to be symmetry-related, their frequency structures are rather different.

Finally, the alternating loop state was revealed to have a hitherto unsuspected structure in which the main modes are locked in a  3:2 frequency ratio.  The Lissajous figure traced by the tip of the thread in this state has an intriguing and unexpected multi-looped pretzel shape.  The reconstruction of this surprising shape was only made possible by side-view visualization.  This observation demonstrates that even the simplest states, found well below the inertial-gravitational regime of nozzle heights for which thread's higher pendulum modes become active, can have complex multi-frequency structures.

    At present, there is no detailed theoretical understanding of all these states of motion and the bifurcations between them.   The simplicity of their frequency structure suggests that a model containing only a few coupled nonlinear modes within the gravitational regime might suffice to explain the phenomena.



\begin{acknowledgments}
During the preparation of this manuscript, we learned of similar but independent work by Pierre-Thomas Brun, Niel M. Ribe and Basile Audoly using a clever numerical simulation of the centerline motion of the thread~\cite{brun_ribe_audoly_2011}.  We gratefully acknowledge conversations with them and with Eitan Grinspun, Jonathan H. P. Dawes, John R. Lister and Maurice J. Blount.  This work was supported by the Natural Science and Engineering Research Council of Canada.
\end{acknowledgments}

%
%


\begin{thebibliography}{99}
\bibitem{shape_note} For larger nozzle heights, the thread motion becomes more complex and a more detailed fit to the shape of the thread would be necessary to infer its contact point.
\end{thebibliography}
\end{document}